\documentclass[twocolumn,tighten]{aastex63}
\usepackage{amssymb, amsmath,framed}

\usepackage{color}

\usepackage{bm}
\expandafter\ifx\csname package@font\endcsname\relax\else
 \expandafter\expandafter
 \expandafter\usepackage
 \expandafter\expandafter
 \expandafter{\csname package@font\endcsname}
\fi
\def\bq{\begin{equation}}
\def\eq{\end{equation}}
\def\bqy{\begin{eqnarray}}
\def\eqy{\end{eqnarray}}

\def\p{\partial}

\def\p{\partial}

\def\R{\mathbb{R}}

 \def\q#1#2{q^{\hspace{1pt}#1}_{\,,\hspace{1pt}#2}}
 \def\a#1#2{a^{\hspace{1pt}#1}_{\,,\hspace{1pt}#2}}
 \def\d#1#2{\delta^{\hspace{1pt}#1}_{\,,\hspace{1pt}#2}}

\begin{document}
\title{\large{Feasibility of Detecting Interstellar Panspermia in Astrophysical Environments}}

\correspondingauthor{Claudio Grimaldi}
\email{claudio.grimaldi@epfl.ch}

\author{Claudio Grimaldi}
\affiliation{Laboratory of Statistical Biophysics, Ecole Polytechnique F\'ed\'erale de Lausanne - EPFL, 1015 Lausanne, Switzerland}
\affiliation{Centro Studi e Ricerche Enrico Fermi, 00184 Roma, Italy}

\author{Manasvi Lingam}
\affiliation{Department of Aerospace, Physics and Space Sciences, Florida Institute of Technology, Melbourne, FL 32901, USA}

\author{Amedeo Balbi}
\affiliation{Dipartimento di Fisica, Universita` di Roma ``Tor Vergata", 00133 Roma, Italy}

\begin{abstract}
The proposition that life can spread from one planetary system to another (interstellar panspermia) has a long history, but this hypothesis is difficult to test through observations. We develop a mathematical model that takes parameters such as the microbial survival lifetime, the stellar velocity dispersion, and the dispersion of ejecta into account in order to assess the prospects for detecting interstellar panspermia. We show that the correlations between pairs of life-bearing planetary systems (embodied in the pair-distribution function from statistics) may serve as an effective diagnostic of interstellar panspermia, provided that the velocity dispersion of ejecta is greater than the stellar dispersion. We provide heuristic estimates of the model parameters for various astrophysical environments, and conclude that open clusters and globular clusters appear to represent the best targets for assessing the viability of interstellar panspermia.\\
\end{abstract}

\section{Introduction} \label{SecIntro}
The modern conception of ‘lithopanspermia’, i.e.\ the idea that life can travel across planetary systems carried by meteoroids and other minor bodies, dates back to at least the 19th century \citep{Kamminga}, but it has regained serious consideration in scientific discussions \citep{Melosh1988, Wesson2010,Wick2010}. This revival was motivated by a number of developments. First, data concerning the resistance of radiation-tolerant organisms in deep space, as well as experimental tests of hypervelocity impacts, show that it is plausible that some organisms can survive the accidental transfer from an inhabited planet to another location \citep{Horneck2010,Onofri2012,Merino2019}. Second, there is now strong evidence that rock fragments have in fact been exchanged between nearby planets in the Solar System \citep{Nyquist2001} and it is conceivable that similar mechanisms can be even more efficient in more densely packed planetary systems, such as the TRAPPIST-1 system \citep{Lingam2017,Krijt2017}. Finally, the direct observation of at least two objects of interstellar origin transiting the Solar Systems has confirmed that the exchange of material is possible even between different planetary systems \citep{Meech2017,Guzik2020}. 

The actual feasibility of lithopanspermia on interstellar distances has been addressed in the past, and shown to be achievable, at least in principle \citep{Zubrin2001,Wallis2004,Napier2004,Ginsburg2018,Siraj2020}, especially in crowded environments, such as in star-forming clusters \citep{Adams2005,Valtonen2009,Belbruno2012} or in the Galactic bulge \citep{CFL18,Balbi2020a}; furthermore, its probability can be significantly enhanced by interactions with binary systems \citep{Lingam2018}. Although the issue is still debated and, admittedly, rather speculative, there are strong theoretical motivations to explore how a viable lithopanspermia mechanism could impact the distribution of life in the Galaxy, particularly if compared to the case where life originates independently in separate locations. 

There are at least two major implications that warrant a careful analysis of the problem \citep{Balbi2021}. First, lithopanspermia may change the estimated extent of the Galactic Habitable Zone (GHZ), since it can act as a compensating factor with respect to potentially catastrophic events that can eradicate life from planetary surfaces \citep{Balbi2020a}. Second, it can alter the statistical impact and interpretation of future biosignature detections via surveys of nearby planetary systems as shown by \citet{Balbi2020}. 

Therefore, we argue it is timely to develop models that can shed some light on the consequences of interstellar lithopanspermia, assuming it is a practical mechanism. In this Letter, we present a theoretical treatment of the expected statistical distribution of inhabited planets in a lithopanspermia scenario, with a particular focus on its correlation properties. Our work generalizes and quantifies the prior analyses by \citet{Lin2015} and \citet{Lingam2016,Manasvi2016}, which dealt with this topic.

\section{Formalism}\label{Sec:Formalism}
To model the spreading of life over interstellar distances, we suppose that impacts of meteoroids and other minor celestial bodies on rocky planets collectively engender a steady ejection rate of matter escaping the gravitational pull of the host star. This assumption has been employed in several publications on interstellar lithopanspermia. \citet{Melosh2003} predicted, for example, that $\sim 15$ rocks of size $> 10$ cm originating from impacts on inner planets exit the Solar System each year on average. On the other hand, when it comes to micron-sized fragments, they are thought to be expelled at an average rate of $\gtrsim 10^{14}$ per year \citep{Napier2004}. While the ejection rate is likely to have been boosted during an epoch of intense bombardment as a result of the higher impact rates \citep{Adams2005,Belbruno2012}, our model does not necessitate knowledge of the exact value of this parameter, as long as it is not infinitesimally small.

A fraction of the rocks ejected from planets harboring a biosphere may encapsulate simple forms of life which, provided that they are appropriately shielded from UV and ionizing radiation as well as desiccation and other extremes, can possibly survive in space over timescales of millions of years \citep{Mil2000,LL21}. On a related note, as per some (rather controversial) studies, Earth-based microbes might be capable of survival for intervals as high as $\sim 100$ Myr under suitable conditions \citep{Cano1995,Vree2000,Morono2020}. 

A world harboring a biosphere could thus be envisioned as being surrounded by a spherical region of radius $R$ sparsely populated by life-bearing ejecta \citep{Wallis2004}. Under favorable circumstances, these objects could potentially seed life on an initially sterile planet located within such spheres of influence, each of which is termed a \emph{Lebenssph{\"a}re} (life-sphere). We roughly estimate the average radius of a Lebenssph{\"a}re by $R=\langle v_o\rangle\tau_o$, where $\tau_o$ is the mean lifetime of the microbial populations encapsulated in the ejecta and $\langle v_o\rangle$ is the mean velocity of the objects in the local frame of the Lebenssph{\"a}re. By adopting a Maxwellian velocity distribution with dispersion $\sigma_o$, the average radius can be expressed as follows:
\begin{equation}
\label{R1}
R=\sqrt{8/\pi}\sigma_o\tau_o\simeq\textrm{5.3 ly}\left(\frac{\sigma_o}{\textrm{1 km s$^{-1}$}}\right)\left(\frac{\tau_o}{\textrm{1 Myr}}\right),
\end{equation}
yielding, for example, $R\simeq 50$-$100$ ly in the event $\sigma_o=10$-$20$ km s$^{-1}$ and $\tau_o$ as large as $1$ Myr. A sphere of influence of this size located in the solar neighborhood would contain $\sim 10^3$-$10^4$ stars and a comparable number of rocky planets, while it would accommodate a number of stars up to four orders of magnitude greater if located in the Galactic bulge. We further note that a seeding sphere of $\sim 50$ ly in radius is much smaller than most typical length scales in the Galaxy. This allows us to consider volume samples of the Galaxy of linear size larger than $R$ but, at the same time, sufficiently small that the distribution of stars in the sample may be treated as homogeneous and isotropic.

If a planet to which life has been successfully transferred through lithopanspermia can sustain an enduring biosphere, it will after some time  develop a new Lebenssph{\"a}re surrounding it, which will potentially seed life on other planets in turn. We make the ostensibly reasonable hypothesis that this mechanism engenders Lebenssph{\"a}ren with a constant birthrate per planet, denoted by $\tau^{-1}$, whereas Lebenssph{\"a}ren formed by spontanous and independent abiogenesis events grow at a constant rate $\gamma$ per unit volume. Furthermore, we assume that the proliferation of life-harboring planets is counterbalanced by random sterilizing events such as nearby supernovas or other cataclysms \citep{GM18}, so that within a sample volume there exists a constant number density $\rho$ of Lebenssph{\"a}ren formed around planets on which life arose spontaneously or by lithopanspermia. As shown in Appendix \ref{AppA}, such steady-state regimes are reached as long as the inequality $L/\tau < 1$ holds true, where $L$ denotes the average persistence lifetime of a Lebenssph{\"a}re, consequently yielding $\rho=\gamma L/(1-\tau/L)$. 

\begin{figure*}
	\begin{center}
			\includegraphics[width=15 cm,clip=true]{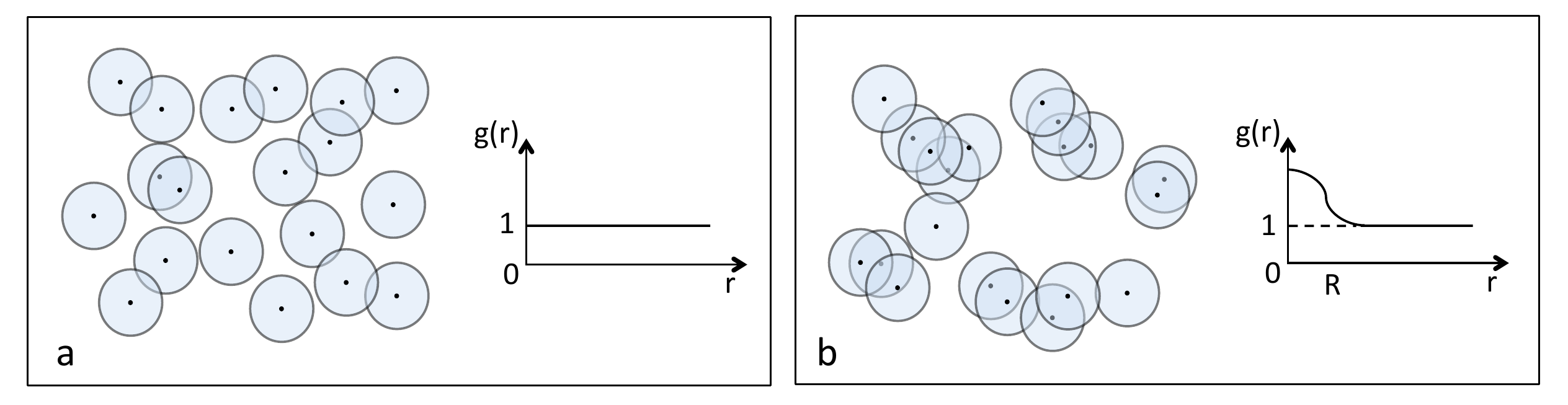}
\caption{Schematic illustration of Lebenssph{\"a}ren (circles) centered around life-bearing planets (points). Case \textbf{a}: Independent abiogenesis results in a uniform spatial distribution of Lebenssph{\"a}ren. Their centers are spatially uncorrelated and the pair distribution function is $g(r)=1$. Case \textbf{b}: When panspermia sets in and the stellar velocities are minimal, the Lebenssph{\"a}ren are aggregated in clusters formed by life-bearing planets at pair distances smaller than $R$. In this scenario, $g(r)$ peaks at $r\lesssim R$.}\label{fig1}
	\end{center}
\end{figure*}

In the following statistical analysis, it is mathematically convenient to first evaluate the static limit where all the stars in the sample have fixed positions that do not change over time; this is not a bad assumption as long as the stellar velocity dispersion is sufficiently low. The Lebenssph{\"a}ren, which we assume all possess the same mean radius $R$ for simplicity, will therefore have their center at given immutable positions. The average number of the centers of Lebenssph{\"a}ren within a distance $R$ from a randomly chosen seeding world is:
\begin{equation}
\label{N1}
N(R)=4\pi\rho\int_0^R\! dr\,r^2g(r)
\end{equation}
where $g(r)$ represents the pair-distribution function defined in such a way that $\rho^2g(r)$ leads to the probability distribution function associated with two Lebenssph{\"a}re centers being situated at a distance $r$ from each other.
If we suppose that life arose independently on different planetary systems (i.e., independent abiogenesis), the Lebenssph{\"a}re centers would be spatially uncorrelated, as schematically shown in Figure~\ref{fig1}a. In this case, it follows that $g(r)=1$ and therefore
\begin{equation}
\label{N2}
N(R)=\frac{4}{3}\pi\rho R^3\equiv\eta.
\end{equation}
We now suppose that a panspermia mechanism along the lines described previously is able to spread life. In the static limit, a planet fertilized through panspermia (which in turn develops its own Lebenssph{\"a}re) must be located at a distance $r \leq R$ from the center of a Lebenssph{\"a}re, as illustrated in Figure~\ref{fig1}b. The coordination number $N(R)$ must therefore be larger than its value in the absence of panspermia, Eq.~\eqref{N2}, because there now exists, within $R$, a higher probability of finding a planet to which life has been transferred:
\begin{equation}
\label{N3}
N(R)=4\pi\rho\int_0^R\! dr\,r^2g(r)=\chi\eta,
\end{equation}
where $\chi\ge 1$ accounts for the enhanced population of life-bearing planets located within $R$; we hereafter christen $\chi$ the panspermia amplification factor (PAF). Equation \eqref{N3} simply states that, compared to the case of spontaneous and independent abiogenesis, the pair-distribution function is accordingly enhanced in the neighborhood of a seeding planet. We specify $g(r)=\chi$ for $r\leq R$, which amounts to assuming that the probability of seeding a planet is uniform for $r\leq R$. Although a more realistic ansatz is feasible, we wish to primarily explicate the qualitative features of this model.

\begin{figure}
	\begin{center}
			\includegraphics[width=\columnwidth,clip=true]{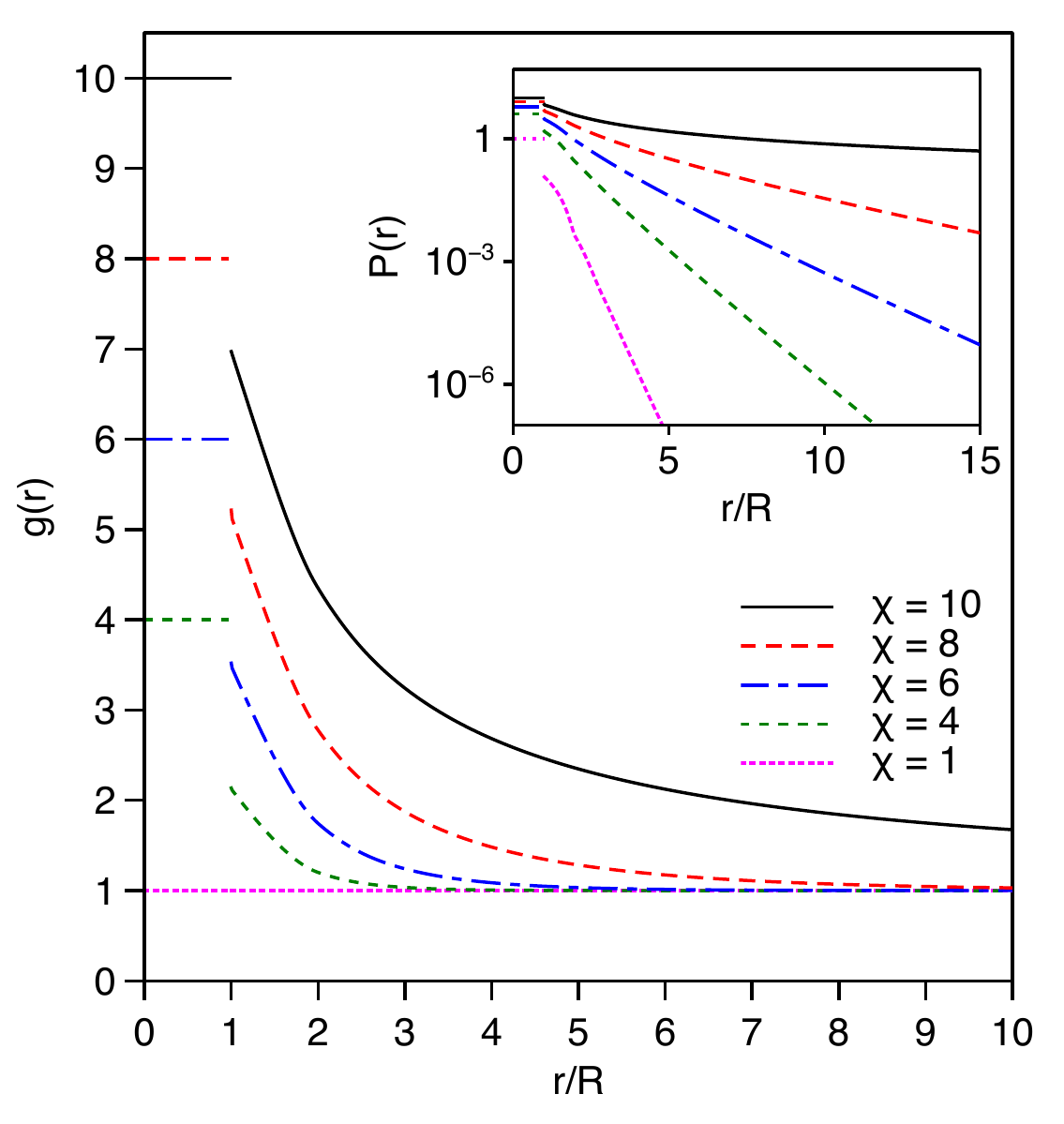}
\caption{Pair-distribution function $g(r)$ calculated in the static limit for $\eta=0.4$ and different values of the panspermia intensity parameter $\chi$. Inset: log-linear plot of the pair-connectedness function demonstrating that $P(r)$ decays exponentially with $r$ when the system is below the percolation threshold $\eta\chi=4$. At the critical point, $P(r)\propto 1/r$.}\label{fig2}
	\end{center}
\end{figure}

At greater distances, we expect the pair-distribution function to vary on a length scale of order $\xi$, eventually reaching the uniform limit $g(r)=1$ for $r\gg\xi$, where $\xi$ measures the typical size of clusters formed by life-bearing planets at pair distances smaller than $R$. 
The spatial distribution of such ``connected'' planets is given by the so-called pair-connectedness function $P(r)$, defined such that $\rho^2P(r)$ gives the probability distribution function associated with finding two Lebenssph{\"a}re centers separated by a distance $r$ and belonging to the same cluster \citep{Torquato2002}. From this definition it follows that $P(r)=g(r)$ for $r\leq R$, while $P(r)$ decays exponentially over distances larger than $\xi$, when no giant cluster of connected planets emerges; in the language of percolation theory, this regime amounts to being below the percolation threshold \citep{Stauffer1994}.

The above considerations suggest the following ansatz:
\begin{equation}
\label{ansatz}
g(r)=1+\frac{\chi-1}{\chi}P(r),
\end{equation}
from which we recover the equality $g(r)=\chi\geq 1$ for $r\leq R$, which we had posited earlier. To find $g(r)$ for $r>R$ we calculate $P(r)$ by following standard techniques developed in continuum percolation theory and briefly described in Appendix \ref{AppB}. Figure~\ref{fig2} depicts $g(r)$ calculated for $\eta=0.4$ and for values of the PAF ranging from $\chi=1$ (no panspermia) up to $\chi=10$, at which point the system reaches the percolation threshold $\eta_c \chi=4$ (Appendix \ref{AppB2}). Aside from the enhanced $g(r)$ for $r\leq R$, we see from Figure \ref{fig2} that the higher is the PAF (i.e., larger $\chi$), the greater is the distance above which $g(r) \rightarrow 1$ is applicable.

This can be understood by noticing that the correlation length scales as $\xi \propto \eta\chi/(1-\eta\chi)$; see Eq. \eqref{xi2} of Appendix \ref{AppB3}, which increases monotonically with $\chi$ and eventually diverges at the percolation threshold of $\eta\chi = 4$. At this point, $P(r)$, and therefore $g(r)-1$ due to Eq. \eqref{ansatz}, decays as a power-law (inset of Figure \ref{fig2}). It is worth pointing out that our panspermia model predicts a critical density of percolation that is inversely proportional to $\chi$ -- specifically, $\eta_c=4/\chi$
as shown in Appendix \ref{AppB2} -- meaning that a sufficiently high PAF is capable of inducing an entire sample-spanning transfer of life even if life-bearing planets are, overall, very rare.

The ansatz \eqref{ansatz} is strictly applicable only to the subcritical regime wherein $\eta < \eta_c$. Once we exceed the percolation threshold, characterized by the formation of a giant cluster \citep{Stauffer1994}, we expect that the pair distribution function generated by the panspermia process would gradually become less peaked as $\eta$ increases beyond the percolation threshold, eventually tending toward $g(r)=1$ for $\eta\gg 1/\chi$. In this limit, indeed, the Lebenssph\"{a}ren would essentially cover the entire sample volume in a uniform fashion.

\begin{figure}
	\begin{center}
			\includegraphics[width=\columnwidth,clip=true]{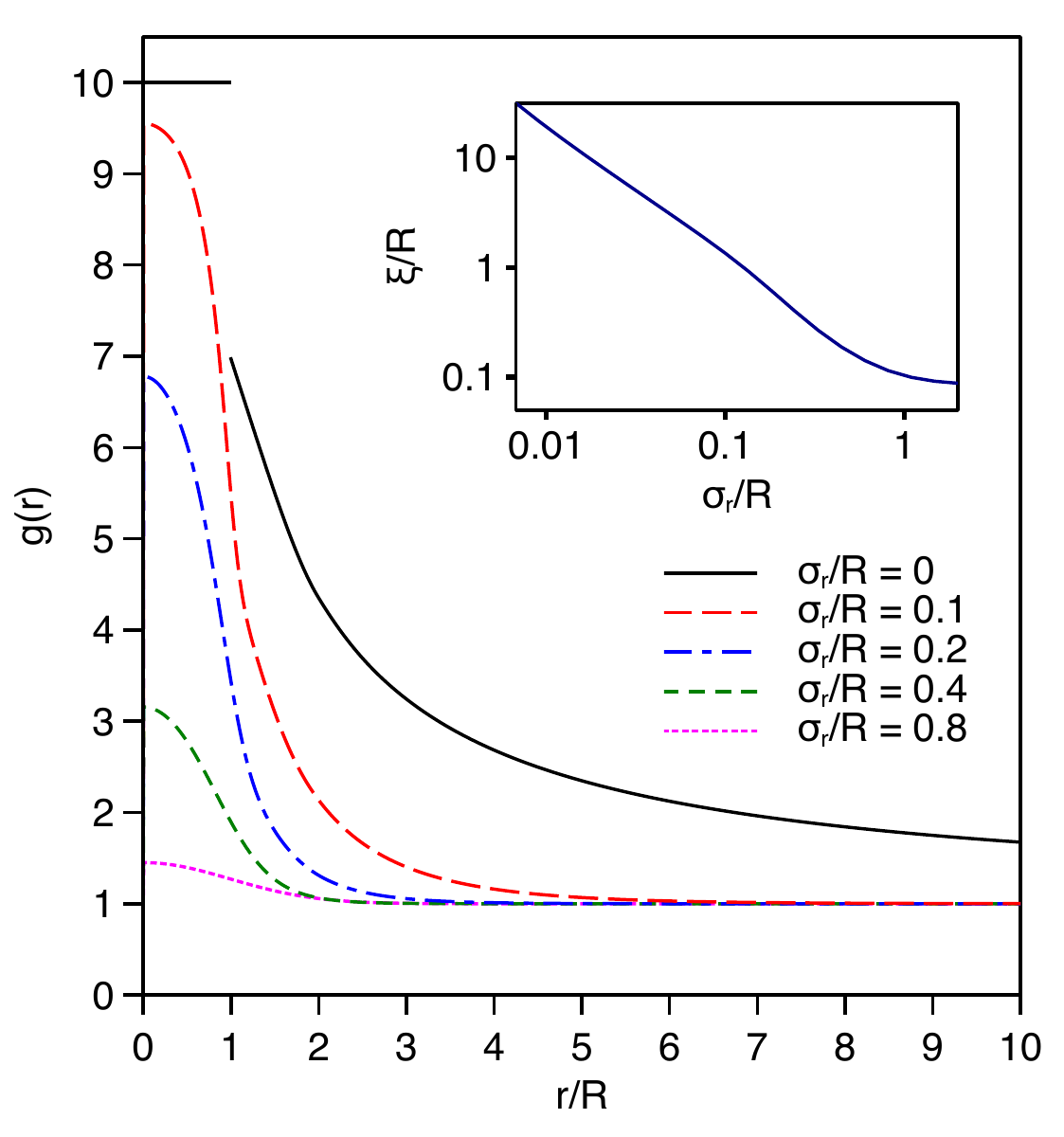}
\caption{Pair-distribution function $g(r)$ calculated for $\eta=0.4$, $\chi=10$ in the case of non-zero stellar velocity dispersion $\sigma_s$. The parameter $\sigma_r=\sqrt{2}\sigma_s\tau$ measures the dispersion of the relative separation of Lebenssph{\"a}re centers; note that $\sigma_r=0$ corresponds to the static limit. As $\sigma_r$ increases, the discontinuity of $g(r)$ at $r=R$ is smeared out and the peak at $r\lesssim R$ is weakened. For $\sigma_r\gtrsim R$ the pair distribution function becomes flat as in the uncorrelated case. Inset: Correlation length $\xi$ plotted as a function of $\sigma_r/R$.}\label{fig3}
	\end{center}
\end{figure}


We will now extend the static panspermia mechanism discussed above to the more realistic case in which stars hosting planetary systems have nonzero relative velocities. Intuitively, even if the static panspermia model may evince pronounced spatial correlations, we would expect them to be nevertheless weakened by the accompanying stellar motion. To explore this effect, we note that for Lebenssph{\"a}ren that are generated via panspermia at a constant rate $\tau^{-1}$, 
the position of a newly formed Lebenssph{\"a}re relative to that of the seeding planet will be translated by $\sim\bm{v}\tau$, where $\bm{v}$ is the relative velocity of 
the two planetary systems. We capture this effect by rewriting the ansatz \eqref{ansatz} as:
\begin{equation}
\label{g2}
g(r)=1+\frac{\chi-1}{\chi}\int\!d\bm{v}\,f(\bm{v})P(\vert\bm{r}+\bm{v}\tau\vert),
\end{equation}
where $f(\bm{v})$ is the distribution function of the relative velocities. Assuming that in the local frame of the sample volume the stars move in random directions with velocities obeying a Maxwell distribution with dispersion $\sigma_s$, Eq.~\eqref{g2} can be recast in the following form:
\begin{equation}
\label{g3}
g(r)=1+\frac{\chi-1}{\chi}\!\int\!d\bm{r}'\,\frac{1}{(2\pi\sigma_r^2)^{3/2}}\exp\!\left(\!-\dfrac{\vert\bm{r}-\bm{r}'\vert^2}{2\sigma_r^2}\right)P(r'),
\end{equation}
where $\sigma_r=\sqrt{2}\sigma_s\tau$ measures the dispersion of the relative distance between two centers of Lebenssph{\"a}ren. Clearly, for $\sigma_r\rightarrow 0$ we recover the ansatz \eqref{ansatz} while for $\sigma_r\gg R$ the pair-distribution of the Lebenssph{\"a}re centers reaches the uniform limit as $g(r) \rightarrow 1+\mathcal{O}(R^3/\sigma_r^3)$. 

We show the crossover from the correlated to the uncorrelated regime in Figure~\ref{fig3}, where we report the numerical calculations of $g(r)$ obtained by self-consistently solving Eq.~\eqref{g3} and the integral equation of $P(r)$ (Appendix \ref{AppB1}). The enhancement of $g(r)$ for $r\leq R$
is rapidly weakened as $\sigma_r$ increases and, simultaneously, the correlation length declines towards the uncorrelated limit (inset of Figure~\ref{fig3}). In fact, even for $\sigma_r/R\approx 1$ the pair-distribution function is essentially indistinguishable from that of a fully uncorrelated system; the latter refers to the case with $\chi=1$ in Figure~\ref{fig2}. 

For $\sigma_r/R\gtrsim 1$, therefore, the Lebenssph{\"a}re centers are uniformly distributed, thereby hindering our ability to infer any information about the extent and intensity of panspermia events if one were to simply measure the distribution of life-bearing planets in a given sample volume through an appropriate survey. 

\section{Discussion}

\begin{table*}
\begin{minipage}{165mm}
\caption{Heuristic estimates of model parameters for various astrophysical environments}
\label{TableSystem}
\vspace{0.1 in}
\begin{tabular}{|c|c|c|c|c|c|c|c|c|c|}
\hline
{\bf System} & $\sigma_o$ & $\sigma_s$ & $\tau_o$ & $\tau$ & $R$ & $\sigma_r$ & $\sigma_r/R$ & $\rho_\star$ & $\zeta_c$ \tabularnewline
\hline
\hline
Solar neighborhood & $10$ & $20$ & $1$ & $1$ & $53$ & $94$ & $1.8$ & $2.9 \times 10^{-3}$ & $2.2 \times 10^{-3}$\tabularnewline
\hline
Globular clusters & $10$ & $5$ & $1$ & $1$ & $53$ & $23.5$ & $0.4$ & $2.9$ & $2.2 \times 10^{-6}$\tabularnewline
\hline
Open clusters & $10$ & $1$ & $1$ & $1$ & $53$ & $4.7$ & $9 \times 10^{-2}$ & $0.3$ & $2.2 \times 10^{-5}$\tabularnewline
\hline
Galactic bulge & $10$ & $100$ & $1$ & $1$ & $53$ & $470$ & $8.9$ & $0.3$ & $2.2 \times 10^{-5}$\tabularnewline
\hline
Galactic halo & $10$ & $150$ & $1$ & $1$ & $53$ & $705$ & $13.3$ & $1.4 \times 10^{-5}$ & $4.6 \times 10^{-1}$\tabularnewline
\hline
\end{tabular}

\medskip

\textbf{\textit{Additional notes:}} $\sigma_o$ and $\sigma_s$ are expressed in units of km s$^{-1}$, $R$ and $\sigma_r$ in units of light years (ly), $\tau$ and $\tau_0$ in units of Myr, and $\rho_\star$ in units of stars per cubic ly. The parameters for the above systems were chosen based on the following references: (i) solar neighborhood \citep{HNA09}, (ii) globular clusters \citep{Baum18}, (iii) open clusters \citep{Valtonen2009,Foster15}, (iv) Galactic bulge \citep{Zhu2017,Balbi2020a}, and (v) Galactic halo \citep{Helmi2008}.
\end{minipage}
\end{table*}

It is now instructive to carry out some fiducial estimates for select astrophysical systems, and consequently assess the efficacy of panspermia. Of the various parameters at play, one of the most ambiguous among them is $\tau$, which is the time elapsed from a seeding
event to the formation of a Lebenssph{\"a}re of radius $R$, owing to which it cannot be smaller than $\tau_o$. If we suppose that life could germinate quickly after the seeding event and form a biosphere, given habitable conditions, it is conceivable that $\tau$ is of the same order as $\tau_o$, owing to which we will employ the condition $\tau \gtrsim \tau_o$.

One other crucial variable is $\rho$ because it regulates $\eta$, as seen from Eq.~\eqref{N2}. Given that $\rho$ is the number density of Lebenssph{\"a}ren, this is not an easy quantity to gauge since it requires us to know the fraction of planetary systems with habitable worlds on which life blossomed into a biosphere ($\zeta_\star$); in essence, therefore, it corresponds to two weakly constrained factors in the Drake equation \citep{Drake65}. As per our postulate, we have $\rho = \zeta_\star \rho_\star$, where $\rho_\star$ is the stellar density. In the static limit, we have shown that the percolation threshold is $\eta_c = 4/\chi$ in Appendix \ref{AppB2}, which implies that $\eta_c < 4$ because $\chi > 1$ by construction. It is therefore feasible to determine an upper bound on $\zeta_\star$ (denoted by $\zeta_c$) in order for $\eta$ to stay below the percolation threshold and in the subcritical regime from Eq.~\eqref{N2} as follows:
\begin{equation}
 \zeta_c = \frac{3}{\pi \rho_\star R^3} 
\end{equation}
If $\rho_\star$ and/or $R$ are sufficiently small, then $\zeta_c > 1$ is mathematically possible, but not physically realizable because we have defined $\zeta$ such that it is smaller than unity. Hence, environments with $\zeta_c < 1$ are liable to always exist in the subcritical regime with $\eta < \eta_c$. Estimating the required value of $\zeta_c$ is helpful because it functions as a heuristic signpost for determining whether we are below or above the percolation threshold. In the latter scenario, as explained in Sec. \ref{Sec:Formalism}, the panspermia correlations manifested in the pair-distribution function could become indistinguishable if $\eta$ is high. 

At this juncture, we emphasize that some of our ensuing results are not sensitive to $\eta$. For instance, the crucial ratio of $\sigma_r/R$ is expressible as
\begin{equation}\label{Sigmaratio}
    \frac{\sigma_r}{R} = \frac{\sqrt{\pi}}{2} \left(\frac{\sigma_s}{\sigma_o}\right) \left(\frac{\tau}{\tau_o}\right),
\end{equation}
which implies that a sufficient condition for $\sigma_r/R\gtrsim 1$ to hold true is $\sigma_s \gtrsim 1.13\,\sigma_o$; this relation follows after utilizing $\tau \gtrsim \tau_o$ from the preceding paragraph. Thus, broadly speaking, if the stellar velocity dispersion is higher than that of the ejecta, the panspermia correlations are effectively washed out. If the opposite is true, then the correlated regime might be manifested, enabling us to discern panspermia through observations, but only provided that $\tau$ is roughly comparable to $\tau_o$.

In what follows, we will suppose that the characteristic dispersion of ejecta from planetary systems is $\sigma_o \sim 10$ km s$^{-1}$, which is close to the mean value of $6.2 \pm 2.7$ km s$^{-1}$ for ejection speeds estimated in \citet{Adams2005}. However, it should be noted that a small fraction of ejecta may have speeds of order $0.1$ km s$^{-1}$ \citep{Belbruno2012}. The estimate for $\tau_o$ is not well understood, since it depends on the size of the object, among other factors, but survival timescales of $\gtrsim 1$ Myr are possible for microbial populations in ejecta with sizes of $\gtrsim 1$ m \citep{Mil2000,Valtonen2009,LL21} and intervals of $\sim 10$-$100$ Myr are not impossible \citep{Cano1995,Morono2020}; we will therefore specify $\tau_0 \sim 1$ Myr hereafter. Although we adopt this fiducial value, our results regarding the viability of detecting panspermia are sensitive to the ratio $\tau/\tau_o$, as seen Eq.~\eqref{Sigmaratio}, and are consequently not directly dependent on the magnitude of $\tau_o$.

We have presented heuristic estimates for the salient parameters of our model in Table \ref{TableSystem}. As $\tau$ is subject to significant uncertainty, we have considered the limiting case of $\tau \sim \tau_o$ herein and adopted the values delineated in the prior discussion. There are two broad inferences that can be drawn from this table:
\begin{enumerate}
    \item We find that $\zeta_c \ll 1$ is valid for all astrophysical environments except the Galactic halo. Thus, insofar as the static limit is concerned, the subcritical regime (with $\eta$ below the percolation threshold) is feasible when the density of Lebenssph{\"a}ren is very low. To put it differently, if only a very small fraction of planetary systems develop Lebenssph{\"a}ren in these astrophysical settings, the subcritical regime is valid.
    \item It was argued earlier that panspermia correlations are discernible only when $\sigma_r/R \lesssim 1$. We notice from Table \ref{TableSystem} that this criterion is comfortably satisfied only in the case of open stellar clusters, although globular clusters and the solar neighborhood are not far removed from this desired limit. 
\end{enumerate}

\section{Conclusions}
We presented the results of a mathematical model describing the dissemination of life over interstellar distances through lithopanspermia. The model depends on a number of parameters that are known to varying degrees of accuracy, but all of them can be empirically constrained in principle.

We have focused on the predicted correlation properties of life-bearing planetary systems.  Our calculations show that an active panspermia process could lead to a distinct amplification of the population of life-bearing planets within a certain characteristic distance compared to the case of independent abiogenesis. This correlation distance is sensitive to the details of the lithopanspermia mechanism and is therefore capable of serving as an observational diagnostic to constrain various scenarios. However, we also demonstrated that the correlations can become attenuated, or even nullified altogether, depending on the astrophysical environment under investigation. As this attenuation is related to dynamical parameters, in particular to the velocity dispersion of stellar systems, it may be predicted to an extent in specific settings.

Hence, based on our formalism, we found that stellar clusters are more promising insofar as detecting the instantiation of panspermia is concerned, although it might still be discernible in our solar neighborhood. On the other hand, crowded environments endowed with high stellar dispersions, such as the galactic bulge, could be so effective at spreading life through lithopanspermia that they are essentially indistinguishable from the case with minimal panspermia and a high abiogenesis rate; in other words, the correlations would be washed out.

Given our findings, at least two further directions are worthy of pursuit in future investigations. First, our work can be expanded and refined by carrying out detailed numerical simulations and proceeding beyond some of the simplifying assumptions adopted, most notably the criterion of being below the percolation limit. Such numerical simulations are expected to yield further insights concerning the feasibility of distinguishing between independent abiogenesis and panspermia in a given astrophysical system, and what number of inhabited worlds need to be detected toward this end. From an observational standpoint, it has been suggested that confirming panspermia at $5\sigma$ confidence requires the sampling of $> 25$ life-bearing worlds in the optimal scenario \citep{Lin2015}.

It is necessary, in the same vein, to investigate the relative weights of proliferation versus sterilization as these process govern the prevalence of life in planetary systems, as shown in Appendix \ref{AppA}. Second, as a separate line of inquiry, experimental studies of the viability of panspermia would pave the way for estimating some of the parameters involved (e.g., survival time of extremophile populations in space), consequently evaluating the prospects of detecting life on other planets and gauging the feasibility of interstellar panspermia.

\acknowledgments
The authors thank the reviewer for the constructive feedback, which was helpful for improving the paper. A.B. was partially funded by the Italian Space Agency through the Life in Space project (ASI N. 2019-3-U.0) and by grant number FQXi-MGA-1801 and FQXi-MGB-1924 from the Foundational Questions Institute and Fetzer Franklin Fund, a donor advised fund of Silicon Valley Community Foundation. 

\appendix
\section{Rate equation for the density of Lebenssph{\"a}ren}
\label{AppA}
We make the assumption that within a given volume sample there is a number density $\rho_0$ of planets on which life arose spontaneously and independently. If we denote by $\gamma$ the rate of formation per unit volume of Lebenssph{\"a}ren created by spontaneous abiogenesis and $L$ embodies their typical lifetime, the rate equation for $\rho_0$ reads
\begin{equation}
\label{rate1}
\frac{d\rho_0}{dt}=\gamma-\frac{\rho_0}{L}.
\end{equation}
In the presence of panspermia, however, there will be an additional number density $\rho_p$ contribution arising because life has been transferred and new Lebenssph{\"a}ren have been accordingly formed. Given that life can be transferred from planets on which life arose spontaneously or via prior panspermia events, the corresponding rate equation is
\begin{equation}
\label{rate2}
\frac{d\rho_p}{dt}=\frac{\rho_0+\rho_p}{\tau}-\frac{\rho_p}{L}
\end{equation}
where $\tau^{-1}$ is the formation rate of Lebenssph{\"a}ren (per planet) due to panspermia and in the last term we have presumed that sterilizing events affect $\rho_p$ to effectively the same degree as $\rho_0$. The rate equation for $\rho=\rho_0+\rho_p$ is therefore given by
\begin{equation}
\label{rate3}
\frac{d\rho}{dt}=\gamma+\frac{\rho}{\tau}-\frac{\rho}{L}.
\end{equation}
The above differential equation admits two types of solutions depending on whether the formation rate of Lebenssph{\"a}ren is greater or smaller than the rate of sterilizing events.
The solution for $L/\tau>1$ gives rise to a number density of life-harboring planets which increases exponentially with time; the special limit of $L/\tau = 1$ would result in the number density growing linearly with time. In these cases, the rate of panspermia events is such that all habitable planets within the sample volume eventually develop a biosphere, thence leading to a spatially uniform distribution of Lebenssph{\"a}ren. On the contrary, when it comes to $L/\tau < 1$, the solution of Eq.~\eqref{rate3} for $t\gg L/(1-L/\tau)$ reaches the steady state regime in which $\rho$ is finite and expressible as
\begin{equation}
\label{rate4}
\rho \rightarrow \frac{\gamma L}{1-L/\tau}.
\end{equation}
It is in such a regime of dynamical equilibrium that the spatial distributions of Lebenssph{\"a}ren is anticipated to show non-trivial correlations.

\section{Connectedness model of panspermia}
\label{AppB}
\subsection{Pair-connectedness function}
\label{AppB1}
In our model of panspermia, we define two planets as being ``connected'' if their relative distance is smaller than $R$, namely, the radius of their Lebenssph{\"a}ren. In this fashion, the pair-distribution function can be decomposed into a ``connected'' part and a ``disconnected'' part: $g(r)=P(r)+B(r)$. Here, $P(r)$ is the pair-connectedness function associated with finding two Lebenssph{\"a}ren, with centers separated by a distance $r$, within the same cluster. $B(r)$ is the pair-blocking function associated with finding two Lebenssph{\"a}ren at distance $r$ \emph{not} belonging to the same cluster \citep{Torquato2002}. From these definitions it follows that $P(r\leq R)=g(r)$ and $B(r\leq R)=0$. It is worth pointing out that the ansatz \eqref{ansatz} adopted for the static limit is actually an approximation for $B(r)$; it can be obtained from
from $g(r)=P(r)+B(r)$ by replacing $B(r)$ with $1-P(r)/\chi$, which satisfies the condition $B(r\leq R)=0$.

Since the pair-connectedness function fully defines our model $g(r)$, Eq.~\eqref{ansatz}, we exploit the standard integral equation method of continuum percolation theory to find $P(r)$ via the solution of the Ornstein-Zernike relation \citep{Chiew1983,DeSimone1986}:
\begin{equation}
\label{OZ1}
P(r)=D(r)+\rho\int\! d\bm{r}D(\vert\bm{r}-\bm{r}'\vert)P(r')
\end{equation}
where $\rho$ is the number density of the Lebenssph{\"a}ren and $D(r)$ is the direct-connectedness function which is applicable to Lebenssph{\"a}ren that are directly connected (i.e., the relative distance between their centers is $r\leq R$). Equation \eqref{OZ1} is solvable by imposing a suitable closure relation. Here, we use $D(r)=0$ for $r> R$, which is basically the well-known Percus-Yevick closure relation utilized in the theory of liquids \citep{Hansen2006}. 

We implement the so-called Baxter factorization to solve Eq.~\eqref{OZ1} under the condition 
\begin{equation}
\label{cond}
\begin{array}{ll}
P(r)=g(r), & r\leq R, \\
D(r)=0, & r>R.
\end{array}
\end{equation}
This amounts to decoupling $P(r)$ from $D(r)$ by introducing a new function $q(r)$, having the property $q(r)=0$ for 
$r<0$ and $r>R$, which is related to the pair-connectedness and direct-connectedness functions by:
\begin{equation}
\label{q1}
\left[\hat{q}(k)\hat{q}(-k)\right]^{-1}=1+\rho\hat{P}(k)=\left[1-\rho\hat{D}(k)\right]^{-1}
\end{equation}
where $\hat{P}(k)$ and $\hat{D}(k)$ are the Fourier transforms of $P(r)$ and $D(r)$, respectively, and
\begin{equation}
\label{q2}
\hat{q}(k)=1-2\pi\rho\int_o^R\!dr\,q(r)e^{ikr}.
\end{equation}
The factorization enables us to express $P(r)$ in terms of $q(r)$ [see \citet{Hansen2006} for a thorough derivation]:
\begin{equation}
\label{OZ2}
rP(r)=-q'(r)+2\pi\rho\int_0^R\!dt\,(r-t)P(\vert r-t\vert)q(t),
\end{equation}
where $q'(r)=dq(r)/dr$. The function $q(r)$ can be found by solving Eq.~\eqref{OZ2} for $0<r<R$, since in this range
$P(r)=g(r)$:
\begin{equation}
\label{OZ3}
rg(r)=-q'(r)+2\pi\rho\int_0^R\!dt\,(r-t)g(\vert r-t\vert)q(t),\,\,\,\,0<r<R.
\end{equation}
In the static limit of our panspermia model, the pair-distribution function for $0<r<R$ is simply $g(r)=\chi$, so that 
the solution Eq.~\eqref{OZ3} is of the form $q'(r)=\alpha r+\beta$. Given the boundary condition $q(R)=0$, we find therefore:
\begin{equation}
\label{q3}
q(r)=\frac{1}{2}\alpha(r^2-R^2)+\beta(r-R).
\end{equation}
Substitution of this result into Eq.~\eqref{OZ3} allows us to find the coefficients $\alpha$ and $\beta$:
\begin{equation}
\label{alfabeta}
\alpha=-\chi\frac{1-\eta\chi/4}{(1+\eta\chi/8)^2},\,\,\,\, \beta=-\frac{3R}{16}\frac{\eta\chi^2}{(1+\eta\chi/8)^2},
\end{equation}
where $\eta=\frac{4}{3}\pi\rho R^3$. At this point, $P(r)$ can be calculated by numerically solving Eq.~\eqref{OZ2}
in the range $r>R$.

When we consider the effects of the stellar velocities, we follow essentially the same steps of the static case with, however, the important difference that now $g(r)$ and $P(r)$ must be calculated self-consistently. In practice, we first integrate numerically Eq.~\eqref{g3} using the pair-connectedness function $P(r)$ calculated for the static limit as described above.
The resulting $g(r)$ is inserted into Eq.~\eqref{OZ3} and the integro-differential equation for $q(r)$ is solved numerically.
Next, $P(r)$ is calculated by solving Eq.~\eqref{OZ2} across the entire range of $r$ and inserted back into Eq.~\eqref{g3}. 
We repeat this procedure until convergence is reached.

\subsection{Percolation threshold}
\label{AppB2}
It should be noted that the Ornstein-Zernike relation \eqref{OZ1} applies below the percolation threshold where no giant cluster of connected centers of Lebenssph{\"a}ren exists. The percolation threshold is the point at which the mean cluster size 
\begin{equation}
\label{S1}
S=1+\rho\int\!d\bm{r}P(r)=1+\rho\hat{P}(0)
\end{equation}
diverges. Using Eqs.~\eqref{q1} and \eqref{q2}, $S$ can be expressed in terms of the function $q(r)$:
\begin{equation}
\label{S2}
S=\left[1-2\pi\rho\int_0^R\!dr\, q(r)\right]^{-2}.
\end{equation}
In the static limit, $S$ can be calculated analytically using Eqs.~\eqref{q3} and \eqref{alfabeta}, yielding
$S=(1+\eta\chi/8)^4/(1-\eta\chi/4)^2$, which diverges at the critical point $\eta\chi=4$. 

\subsection{Correlation length}
\label{AppB3}
We calculate the correlation length $\xi$ by solving Eq.~\eqref{OZ2} for $r\gg R$. Letting $f(r)=rP(r)$ we can recast $\eqref{OZ2}$ as a differential equation for $f(r)$:
\begin{equation}
\label{OZ4}
f(r)=2\pi\rho\int_0^R\! dt\, f(r-t)q(t)\simeq 2\pi\rho f(r)\int_0^R\! dt\,q(t)-2\pi\rho f'(r)\int_0^R\! dt\,t q(t),\,\,\,r\gg R,
\end{equation} 
where we have used $f(r-t)\simeq f(r)-f'(r) t$ for $t\leq R\ll r$. It is straightforward to solve Eq.~\eqref{OZ4} to find $P(r)\propto\exp(-r/\xi)/r$, where the correlation length is given by:
\begin{equation}
\label{xi1}  
\xi=\dfrac{2\pi\rho\int_0^R\!dt\,tq(t)}{1-2\pi\rho\int_0^R\!dt\,q(t)}.
\end{equation}
In the static limit, $\xi$ can be calculated analytically by inserting Eqs.~\eqref{q3} and \eqref{alfabeta} into the 
above expression:
\begin{equation}
\label{xi2}
\xi/R=\frac{3}{16}\frac{\eta\chi}{1-\eta\chi/4}.
\end{equation}
Note that the percolation length also diverges at the percolation threshold of $\eta\chi=4$ derived in Appendix \ref{AppB2}, which is a classic feature of percolation theory \citep{Stauffer1994}.

When the stellar motion is taken into consideration, the resulting correlation length is readily obtained by inserting 
into Eq.~\eqref{xi1} the function $q(r)$, which is calculated numerically as described in Appendix \ref{AppB1}.

\bibliographystyle{aasjournal}
\bibliography{PansCor}

\begin{thebibliography}{}
\expandafter\ifx\csname natexlab\endcsname\relax\def\natexlab#1{#1}\fi
\providecommand{\url}[1]{\href{#1}{#1}}
\providecommand{\dodoi}[1]{doi:~\href{http://doi.org/#1}{\nolinkurl{#1}}}
\providecommand{\doeprint}[1]{\href{http://ascl.net/#1}{\nolinkurl{http://ascl.net/#1}}}
\providecommand{\doarXiv}[1]{\href{https://arxiv.org/abs/#1}{\nolinkurl{https://arxiv.org/abs/#1}}}

\bibitem[{{Adams} \& {Spergel}(2005)}]{Adams2005}
{Adams}, F.~C., \& {Spergel}, D.~N. 2005, Astrobiology, 5, 497,
  \dodoi{10.1089/ast.2005.5.497}

\bibitem[{Balbi(2021)}]{Balbi2021}
Balbi, A. 2021, in Planet Formation and Panspermia: New Prospects for the
  Movement of Life through Space, ed. B.~Vukotic, R.~Gordon, \& J.~Seckbach
  (Beverly, MA: Wiley-Scrivener Publishing)

\bibitem[{Balbi \& Grimaldi(2020)}]{Balbi2020}
Balbi, A., \& Grimaldi, C. 2020, Proceedings of the National Academy of
  Sciences, 202007560, \dodoi{10.1073/pnas.2007560117}

\bibitem[{Balbi {et~al.}(2020)Balbi, Hami, \&
  Kova{\v{c}}evi{\'{c}}}]{Balbi2020a}
Balbi, A., Hami, M., \& Kova{\v{c}}evi{\'{c}}, A. 2020, Life, 10, 132,
  \dodoi{10.3390/life10080132}

\bibitem[{{Baumgardt} \& {Hilker}(2018)}]{Baum18}
{Baumgardt}, H., \& {Hilker}, M. 2018, Mon. Not. R. Astron. Soc., 478, 1520,
  \dodoi{10.1093/mnras/sty1057}

\bibitem[{{Belbruno} {et~al.}(2012){Belbruno}, {Moro-Mart{\'\i}n}, {Malhotra},
  \& {Savransky}}]{Belbruno2012}
{Belbruno}, E., {Moro-Mart{\'\i}n}, A., {Malhotra}, R., \& {Savransky}, D.
  2012, Astrobiology, 12, 754, \dodoi{10.1089/ast.2012.0825}

\bibitem[{{Cano} \& {Borucki}(1995)}]{Cano1995}
{Cano}, R.~J., \& {Borucki}, M.~K. 1995, Science, 268, 1060,
  \dodoi{10.1126/science.7538699}

\bibitem[{{Chen} {et~al.}(2018){Chen}, {Forbes}, \& {Loeb}}]{CFL18}
{Chen}, H., {Forbes}, J.~C., \& {Loeb}, A. 2018, Astrophys. J. Lett., 855, L1,
  \dodoi{10.3847/2041-8213/aaab46}

\bibitem[{{Chiew} \& {Glandt}(1983)}]{Chiew1983}
{Chiew}, Y.~C., \& {Glandt}, E.~D. 1983, J. Phys. A: Math. Gen., 16, 2599,
  \dodoi{10.1088/0305-4470/16/11/026}

\bibitem[{{DeSimone} {et~al.}(1986){DeSimone}, {Demoulini}, \&
  {Stratt}}]{DeSimone1986}
{DeSimone}, T., {Demoulini}, S., \& {Stratt}, R.~M. 1986, J. Chem. Phys., 85,
  391, \dodoi{10.1063/1.451615}

\bibitem[{{Drake}(1965)}]{Drake65}
{Drake}, F.~D. 1965, {The Radio Search for Intelligent Extraterrestrial Life},
  ed. G.~{Mamikunian} \& M.~H. {Briggs} (Oxford: Pergamon Press), 323--345

\bibitem[{{Foster} {et~al.}(2015){Foster}, {Cottaar}, {Covey}, {Arce}, {Meyer},
  {Nidever}, {Stassun}, {Tan}, {Chojnowski}, {da Rio}, {Flaherty}, {Rebull},
  {Frinchaboy}, {Majewski}, {Skrutskie}, {Wilson}, \& {Zasowski}}]{Foster15}
{Foster}, J.~B., {Cottaar}, M., {Covey}, K.~R., {et~al.} 2015, Astrophys. J.,
  799, 136, \dodoi{10.1088/0004-637X/799/2/136}

\bibitem[{{Ginsburg} {et~al.}(2018){Ginsburg}, {Lingam}, \&
  {Loeb}}]{Ginsburg2018}
{Ginsburg}, I., {Lingam}, M., \& {Loeb}, A. 2018, Astrophys. J. Lett., 868,
  L12, \dodoi{10.3847/2041-8213/aaef2d}

\bibitem[{{Gowanlock} \& {Morrison}(2018)}]{GM18}
{Gowanlock}, M.~G., \& {Morrison}, I.~S. 2018, Astrobiology: Exploring Life on
  Earth and Beyond, Vol.~1, {The Habitability of our Evolving Galaxy}, ed.
  P.~H. {Rampelotto}, J.~{Seckbach}, \& R.~{Gordon} (Cambridge: Academic
  Press), 149--171, \dodoi{10.1016/B978-0-12-811940-2.00007-1}

\bibitem[{{Guzik} {et~al.}(2020){Guzik}, {Drahus}, {Rusek}, {Waniak},
  {Cannizzaro}, \& {Pastor-Marazuela}}]{Guzik2020}
{Guzik}, P., {Drahus}, M., {Rusek}, K., {et~al.} 2020, Nat. Astron., 4, 53,
  \dodoi{10.1038/s41550-019-0931-8}

\bibitem[{Hansen \& McDonald(2006)}]{Hansen2006}
Hansen, J.-P., \& McDonald, I.~R. 2006, Theory of Simple Liquids (London:
  Elsevier)

\bibitem[{{Helmi}(2008)}]{Helmi2008}
{Helmi}, A. 2008, Astron. Astrophys. Rev., 15, 145,
  \dodoi{10.1007/s00159-008-0009-6}

\bibitem[{{Holmberg} {et~al.}(2009){Holmberg}, {Nordstr{\"o}m}, \&
  {Andersen}}]{HNA09}
{Holmberg}, J., {Nordstr{\"o}m}, B., \& {Andersen}, J. 2009, Astron.
  Astrophys., 501, 941, \dodoi{10.1051/0004-6361/200811191}

\bibitem[{Horneck {et~al.}(2010)Horneck, Klaus, \& Mancinelli}]{Horneck2010}
Horneck, G., Klaus, D.~M., \& Mancinelli, R.~L. 2010, Microbiology and
  Molecular Biology Reviews, 74, 121, \dodoi{10.1128/mmbr.00016-09}

\bibitem[{{Kamminga}(1982)}]{Kamminga}
{Kamminga}, H. 1982, Vistas Astron., 26, 67,
  \dodoi{10.1016/0083-6656(82)90001-0}

\bibitem[{{Krijt} {et~al.}(2017){Krijt}, {Bowling}, {Lyons}, \&
  {Ciesla}}]{Krijt2017}
{Krijt}, S., {Bowling}, T.~J., {Lyons}, R.~J., \& {Ciesla}, F.~J. 2017,
  Astrophys. J. Lett., 839, L21, \dodoi{10.3847/2041-8213/aa6b9f}

\bibitem[{{Lin} \& {Loeb}(2015)}]{Lin2015}
{Lin}, H.~W., \& {Loeb}, A. 2015, Astrophys. J. Lett., 810, L3,
  \dodoi{10.1088/2041-8205/810/1/L3}

\bibitem[{{Lingam}(2016{\natexlab{a}})}]{Lingam2016}
{Lingam}, M. 2016{\natexlab{a}}, Mon. Not. R. Astron. Soc., 455, 2792,
  \dodoi{10.1093/mnras/stv2533}

\bibitem[{{Lingam}(2016{\natexlab{b}})}]{Manasvi2016}
---. 2016{\natexlab{b}}, Astrobiology, 16, 418, \dodoi{10.1089/ast.2015.1411}

\bibitem[{{Lingam} \& {Loeb}(2017)}]{Lingam2017}
{Lingam}, M., \& {Loeb}, A. 2017, Proc. Natl. Acad. Sci. USA, 114, 6689,
  \dodoi{10.1073/pnas.1703517114}

\bibitem[{{Lingam} \& {Loeb}(2018)}]{Lingam2018}
---. 2018, Astron. J., 156, 193, \dodoi{10.3847/1538-3881/aae09a}

\bibitem[{{Lingam} \& {Loeb}(2021)}]{LL21}
---. 2021, {Life in the Cosmos: From Biosignatures to Technosignatures}
  (Cambridge: Harvard University Press).
\newblock \url{https://www.hup.harvard.edu/catalog.php?isbn=9780674987579}

\bibitem[{{Meech} {et~al.}(2017){Meech}, {Weryk}, {Micheli}, {Kleyna},
  {Hainaut}, {Jedicke}, {Wainscoat}, {Chambers}, {Keane}, {Petric}, {Denneau},
  {Magnier}, {Berger}, {Huber}, {Flewelling}, {Waters}, {Schunova-Lilly}, \&
  {Chastel}}]{Meech2017}
{Meech}, K.~J., {Weryk}, R., {Micheli}, M., {et~al.} 2017, Nature, 552, 378,
  \dodoi{10.1038/nature25020}

\bibitem[{{Melosh}(1988)}]{Melosh1988}
{Melosh}, H.~J. 1988, Nature, 332, 687, \dodoi{10.1038/332687a0}

\bibitem[{{Melosh}(2003)}]{Melosh2003}
---. 2003, Astrobiology, 3, 207, \dodoi{10.1089/153110703321632525}

\bibitem[{{Merino} {et~al.}(2019){Merino}, {Aronson}, {Bojanova},
  {Feyhl-Buska}, {Wong}, {Zhang}, \& {Giovannelli}}]{Merino2019}
{Merino}, N., {Aronson}, H.~S., {Bojanova}, D.~P., {et~al.} 2019, Front.
  Microbiol., 10, 780, \dodoi{10.3389/fmicb.2019.00780}

\bibitem[{{Mileikowsky} {et~al.}(2000){Mileikowsky}, {Cucinotta}, {Wilson},
  {Gladman}, {Horneck}, {Lindegren}, {Melosh}, {Rickman}, {Valtonen}, \&
  {Zheng}}]{Mil2000}
{Mileikowsky}, C., {Cucinotta}, F.~A., {Wilson}, J.~W., {et~al.} 2000, Icarus,
  145, 391, \dodoi{10.1006/icar.1999.6317}

\bibitem[{{Morono} {et~al.}(2020){Morono}, {Ito}, {Hoshino}, {Terada}, {Hori},
  {Ikehara}, {D'Hondt}, \& {Inagaki}}]{Morono2020}
{Morono}, Y., {Ito}, M., {Hoshino}, T., {et~al.} 2020, Nat. Commun., 11, 3626,
  \dodoi{10.1038/s41467-020-17330-1}

\bibitem[{{Napier}(2004)}]{Napier2004}
{Napier}, W.~M. 2004, Mon. Not. R. Astron. Soc., 348, 46,
  \dodoi{10.1111/j.1365-2966.2004.07287.x}

\bibitem[{{Nyquist} {et~al.}(2001){Nyquist}, {Bogard}, {Shih}, {Greshake},
  {St{\"o}ffler}, \& {Eugster}}]{Nyquist2001}
{Nyquist}, L.~E., {Bogard}, D.~D., {Shih}, C.~Y., {et~al.} 2001, Space Sci.
  Rev., 96, 105

\bibitem[{{Onofri} {et~al.}(2012){Onofri}, {de la Torre}, {de Vera}, {Ott},
  {Zucconi}, {Selbmann}, {Scalzi}, {Venkateswaran}, {Rabbow}, {S{\'a}nchez
  I{\~n}igo}, \& {Horneck}}]{Onofri2012}
{Onofri}, S., {de la Torre}, R., {de Vera}, J.-P., {et~al.} 2012, Astrobiology,
  12, 508, \dodoi{10.1089/ast.2011.0736}

\bibitem[{{Siraj} \& {Loeb}(2020)}]{Siraj2020}
{Siraj}, A., \& {Loeb}, A. 2020, Life, 10, 44, \dodoi{10.3390/life10040044}

\bibitem[{Stauffer \& Aharony(1994)}]{Stauffer1994}
Stauffer, D., \& Aharony, A. 1994, Introduction to Percolation Theory (London:
  Taylor and Francis)

\bibitem[{Torquato(2002)}]{Torquato2002}
Torquato, S. 2002, Random Heterogeneous Materials: Microstructure and
  Macroscopic Properties (New York: Springer)

\bibitem[{{Valtonen} {et~al.}(2009){Valtonen}, {Nurmi}, {Zheng}, {Cucinotta},
  {Wilson}, {Horneck}, {Lindegren}, {Melosh}, {Rickman}, \&
  {Mileikowsky}}]{Valtonen2009}
{Valtonen}, M., {Nurmi}, P., {Zheng}, J.-Q., {et~al.} 2009, Astrophys. J., 690,
  210, \dodoi{10.1088/0004-637X/690/1/210}

\bibitem[{{Vreeland} {et~al.}(2000){Vreeland}, {Rosenzweig}, \&
  {Powers}}]{Vree2000}
{Vreeland}, R.~H., {Rosenzweig}, W.~D., \& {Powers}, D.~W. 2000, Nature, 407,
  897, \dodoi{10.1038/35038060}

\bibitem[{{Wallis} \& {Wickramasinghe}(2004)}]{Wallis2004}
{Wallis}, M.~K., \& {Wickramasinghe}, N.~C. 2004, Mon. Not. R. Astron. Soc.,
  348, 52, \dodoi{10.1111/j.1365-2966.2004.07355.x}

\bibitem[{{Wesson}(2010)}]{Wesson2010}
{Wesson}, P.~S. 2010, Space Sci. Rev., 156, 239,
  \dodoi{10.1007/s11214-010-9671-x}

\bibitem[{{Wickramasinghe}(2010)}]{Wick2010}
{Wickramasinghe}, C. 2010, Int. J. Astrobiol., 9, 119,
  \dodoi{10.1017/S1473550409990413}

\bibitem[{{Zhu} {et~al.}(2017){Zhu}, {Udalski}, {Novati}, {Chung}, {Jung},
  {Ryu}, {Shin}, {Gould}, {Lee}, {Albrow}, {Yee}, {Han}, {Hwang}, {Cha}, {Kim},
  {Kim}, {Kim}, {Kim}, {Lee}, {Park}, {Pogge}, {KMTNet Collaboration},
  {Poleski}, {Mr{\'o}z}, {Pietrukowicz}, {Skowron}, {Szyma{\'n}ski},
  {KozLowski}, {Ulaczyk}, {Pawlak}, {OGLE Collaboration}, {Beichman}, {Bryden},
  {Carey}, {Fausnaugh}, {Gaudi}, {Henderson}, {Shvartzvald}, {Wibking}, \&
  {Spitzer Team}}]{Zhu2017}
{Zhu}, W., {Udalski}, A., {Novati}, S.~C., {et~al.} 2017, Astron. J., 154, 210,
  \dodoi{10.3847/1538-3881/aa8ef1}

\bibitem[{{Zubrin}(2001)}]{Zubrin2001}
{Zubrin}, R. 2001, J. Br. Interplanet. Soc., 54, 262

\end{thebibliography}

\end{document}